%% file: ionMobility.tex
\documentclass[a4paper,11pt,review,number,sort&compress,superscriptaddress,groupedaddress,altaffilletter,floatfix]{article}
\pdfoutput=1 

\usepackage{jinstpub} 
\usepackage{makecell}
\usepackage{hyperref}
\usepackage{xcolor}

\usepackage{subcaption}

\renewcommand{\thepage}{}
\renewcommand{\thepage}{\arabic{page}}
\makeatletter
\def\l@subsubsection#1#2{}
\renewcommand\@dotsep{10000}
\makeatother

\begin{document}
\input{ionMobility-frontmatter.tex}

\maketitle
\flushbottom

\section{Introduction}
The study of ion migration through diverse media has long been of interest in physics. Measurements of heavy ion migration through noble gases \cite{ref:HeavyIons} have important applications with implications for public health. Specifically, they can be applied towards the measurement of radon contamination in air, the removal of radon progeny by electrostatic precipitation, ion mobility spectrometry, and even earthquake prediction. Measurements of the ion fraction and ion mobility of radon progeny are also of extreme interest to rare physics event searches conducted with noble liquid detectors. The mitigation of critical radon-related backgrounds allows these experiments to achieve higher sensitivity \cite{ref:LUXRn, ref:XENONRn, ref:nEXO, ref:GERDA}.
\par Radon is the bane of all low-background experiments. As a noble gas it is not very chemically reactive. It diffuses to fill whatever volume it enters, and decays by alpha emission to leave behind additional radioactive isotopes that can easily plate out on surfaces. The daughter isotopes in all alpha particle decays are initially negatively charged due to the emission of an alpha particle (+2$e$), however, they can easily lose valence electrons in interactions with the surrounding matter. For each alpha decay there is then some probability that the daughter will ultimately be either neutral or positively charged \cite{ref:RnDaughters}. 
Charged decay products will drift under external electric fields onto surfaces to which they will adhere. It is important to estimate the observed rate of decays of radon and its daughters in experiments such as these because alpha and beta-emitting isotopes present in the bulk of detectors and on surfaces can mimic the signals of interest in searches for WIMP (Weakly Interacting Massive Particle) dark matter, neutrinoless double-beta (0$\nu\beta\beta$) decay, as well as low-energy neutrino experiments. 
\par
\begin{figure}
    \centering
    \includegraphics[width=0.6\linewidth]{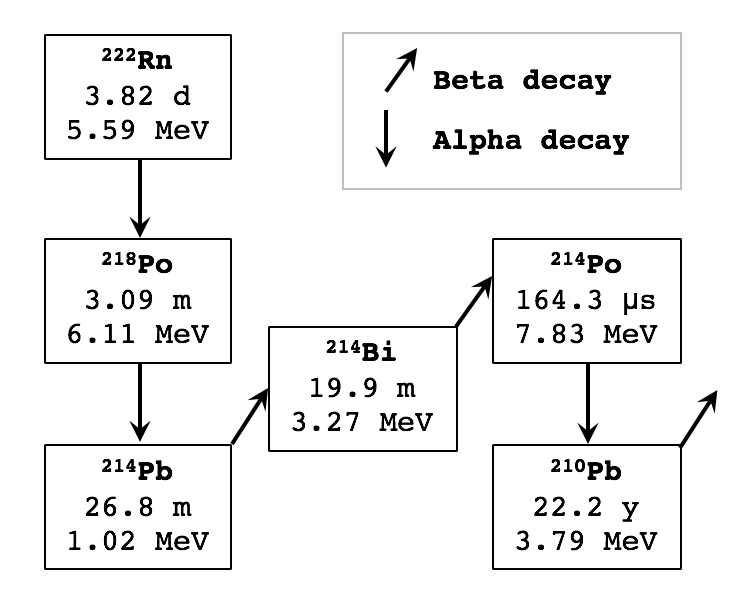}
    \caption{A portion of the uranium-238 series. The Q-value and lifetime is given for each isotope.}
    \label{fig:UChain}
\end{figure}

 Though radon and its daughters form a potentially dangerous background to rare event searches, they also provide an opportunity for detector calibration. Alpha decays can be used as standard candles, and the known lifetimes between decays in the common uranium and thorium chains can be exploited. For example, as shown in Fig.~\ref{fig:UChain}, the alpha decay of $^{222}$Rn is followed by the alpha decay of $^{218}$Po with a half life of 3.09 minutes. These events can be identified by their energies (Q[$^{222}$Rn] = 5.59~MeV, Q[$^{218}$Po] = 6.11~MeV) and associated with each other by their coincidence in time. In this way, the presence of radon and its daughters allows us to measure the characteristics and behavior of specific decay products $in\ situ$, facilitating their proper modelling and mitigation.
\par Knowledge of charged daughter fraction and ion mobility can also influence suppression strategies for other radioactive backgrounds in rare event searches \cite{ref:nEXO, ref:GERDA}. For example, for next generation 0$\nu\beta\beta$ experiments, like nEXO, there is interest in recovering and identifying the $^{136}$Ba daughter produced in the 0$\nu\beta\beta$ decay of $^{136}$Xe in order to eliminate all background except for a small contribution from the two neutrino decay mode (2$\nu\beta\beta$) \cite{ref:BaTagging}. However, this method requires knowledge of the fraction of $^{136}$Ba that is charged, the magnitude of that charge, and how far $^{136}$Ba ions may drift in the detector. Additionally, the GERDA and LEGEND experiments use a liquid argon veto region, and must contend with $^{42}$Ar beta-decaying to $^{42}$K, which can then drift towards the main Ge detectors. Any strategy for suppressing $^{42}$K background events must account for the charged daughter fraction and ion drift. Although $^{218}$Po, $^{136}$Ba, and $^{42}$K have very different masses, and therefore are likely to have different ion drift velocities, studies of radon daughters can provide insight and a framework for future studies with other specific decay products.
\par The EXO-200 collaboration measured the fraction of charged $^{218}$Po and $^{214}$Bi created from alpha and beta decays in liquid xenon (LXe) to be 50.3 $\pm$ 3\% and 76.3 $\pm$ 6.2\%, respectively. They also measured the mobility of $^{218}$Po ions in LXe and observed that the ion velocity decreased as a function of the time the ions spend in motion. The ions had an initial mobility of 0.390 $\pm$ 0.006~cm$^2$/(kV s), which decreased to 0.219 $\pm$ 0.004~cm$^2$/(kV s). The time constant associated with the mobility decrease was proportional to the electron drift-lifetime in EXO-200, so impurities in the LXe are thought to play a role in the mobility variation \cite{ref:EXOMobility}. The effective mobility due to molecular-ion formation in a system consisting of a barium ion surrounded by xenon gas has been considered in the literature \cite{ref:BaClusterIons}. The effective mobility is density-dependent in gas, but the effect in liquid is unclear. The timescale for the formation of molecular-ion structures, measured for Ar$^+$ in Ar gas to be 8.9~ns \cite{ref:ArClusterIons}, is too fast to contribute to the time-dependent mobility observed in EXO-200.  
\par In liquid argon we know of no measurement of the charged daughter fraction of $^{218}$Po produced in the $^{222}$Rn decay, or $^{218}$Po ion mobility.
We present a measurement of the charged daughter fraction for $^{218}$Po produced in the alpha decay of $^{222}$Rn, and the mobility of $^{218}$Po$^+$. Such a measurement is of particular interest to the future argon dark matter program, DarkSide-20k, and may be relevant to the ongoing neutrino physics program using large liquid argon time projection chambers \cite{ref:SBN, ref:DUNE}. 
\par In Sec. \ref{sec:apparatus} we describe the apparatus and available data, followed by a description of the selection of $^{222}$Rn-$^{218}$Po (radon-polonium) coincidences in Sec. \ref{sec:EventSel}. In Sec.~\ref{sec:RnPoDataVel} we extract the charged daughter fraction and ion mobility from the velocity distribution of $^{218}$Po. In Sec. \ref{sec:RnPoSim} we model radon-polonium coincidences, both for a generic detector setup and the restricted analysis volume used in this work. Lastly, in Sec.~\ref{sec:RnPoTDecayFits} we fit the decay time spectrum for identified radon-polonium pairs in DarkSide-50 data with the derived model. 

\section{The DarkSide-50 detector and available data}
\label{sec:apparatus}
\begin{figure}
    \centering
    \includegraphics[width=0.8\linewidth]{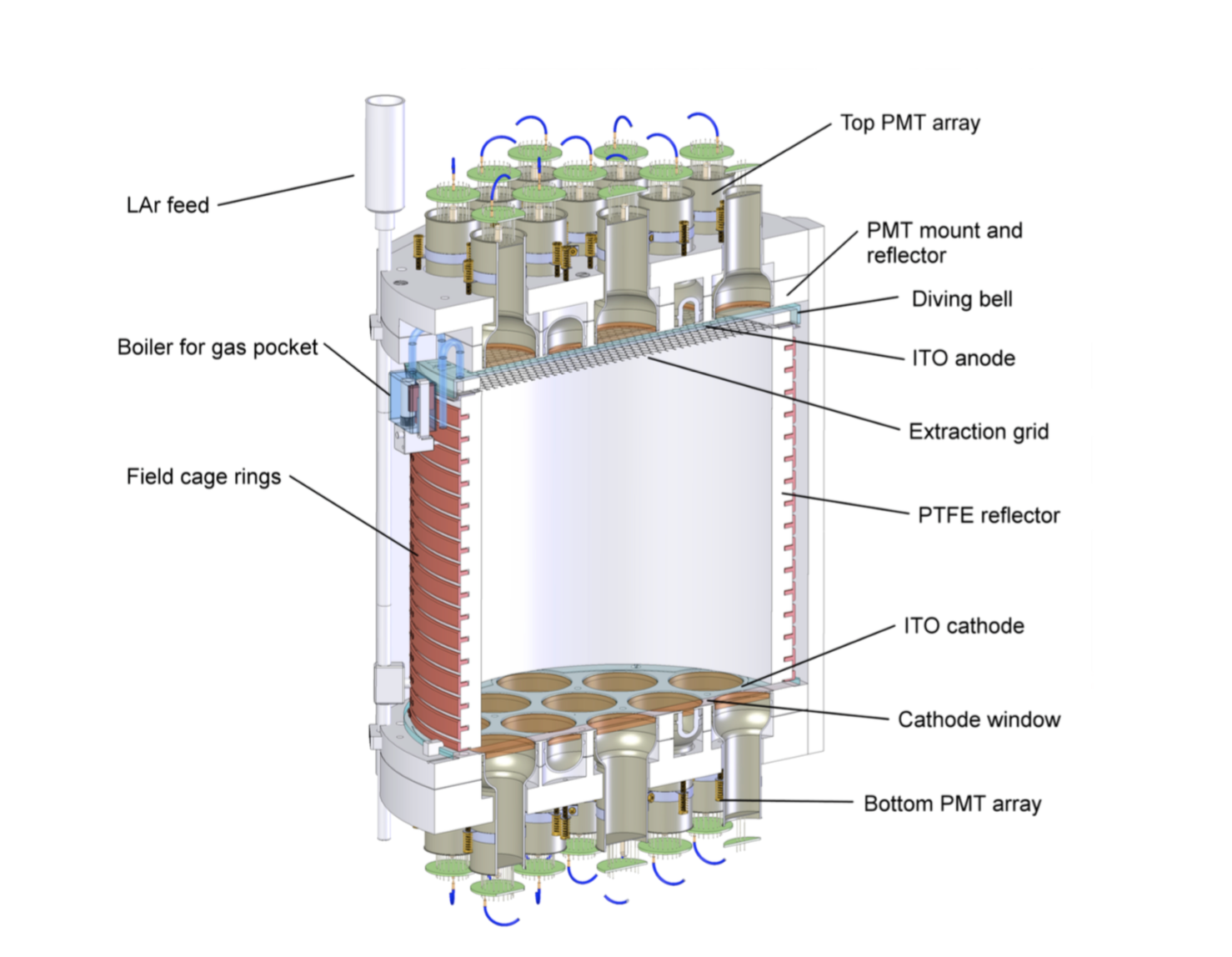}
    \caption{A schematic of the DarkSide-50 TPC.}
    \label{fig:TPC}
\end{figure}

DarkSide-50 is the current phase of the DarkSide WIMP dark matter search program, operating underground at the Laboratori Nazionali del Gran Sasso in Italy. The detector is a dual-phase argon Time Projection Chamber (TPC), housed within a veto system of liquid scintillator and water Cherenkov detectors. The cylindrical TPC volume is filled with liquid argon (LAr), with a thin layer of gaseous argon (GAr) at the top. The TPC is observed through fused-silica windows by two arrays of 19 3" Hamamatsu R11065 photomultiplier tubes (PMTs), one at the top and one at the bottom of the detector. The windows are coated with Indium-Tin-Oxide (ITO), which acts as the cathode (bottom) and anode (top) of the TPC. During normal WIMP search running the 50 kg LAr bulk is subject to a uniform 200 V/cm drift electric field oriented such that negative charges drift towards the gas layer, and positive charges towards the cathode. Interactions in the liquid argon produce excitation and ionization. The de-excitation of argon dimers produces prompt scintillation called S1. Non-recombined ionization electrons are drifted by the electric field to the gas layer, where they produce a secondary scintillation signal, S2, by electroluminescence. The S2 signal provides 3D position information. Depth within the detector ($d$) is given by the drift time of electrons from the interaction site to the gas layer, observable as the time separation between S1 and S2. Transverse position is given by the light distribution over the PMTs. A drawing of the DarkSide-50 TPC can be found in Fig. \ref{fig:TPC}, and full details of the detector can be found in Ref. \cite{ref:50d}.
\par This work uses a dataset from a 532 live-day exposure with underground argon (UAr), which is substantially reduced in $^{39}$Ar activity \cite{ref:70d}. This data was acquired between August 2, 2015, and October 4, 2017 and had an average trigger rate of 1.5 Hz. This is the same dataset used to perform the fully blinded WIMP analysis in Ref. \cite{ref:532d}, as well as a recent S2-only analysis probing low-mass dark matter \cite{ref:S2Only}. 

\section{Event Selection and the Radon-Polonium Search Algorithm}
\label{sec:EventSel}
For this search we are looking for two alpha decays captured in separate event windows (meaning two distinct events with separately triggered acquisition windows). In principle the $^{222}$Rn and $^{218}$Po decays can occur close enough in time that they inhabit the same event window (440~$\mu$s duration), or one decay may be lost to the cut on the time-since-previous-event (1.35~ms duration) that is part of our standard analysis. The lifetime of $^{218}$Po is 268.2~s \cite{ref:Tau}, therefore the total probability of the two decays occurring within the same window or lost to the discussed cut is $P = \frac{1}{268.2 \text{s}} \int_0^{0.00179 s}e^{\frac{-t}{268.2 s}} = 7 \times 10^{-6}$, which is negligible.  The trigger rate during underground argon running in DarkSide-50 is $\sim$1.5~Hz, so we expect other independent events to occur between the alpha decays of interest. 
\par
\begin{figure}
\centering
  \includegraphics[width=\linewidth]{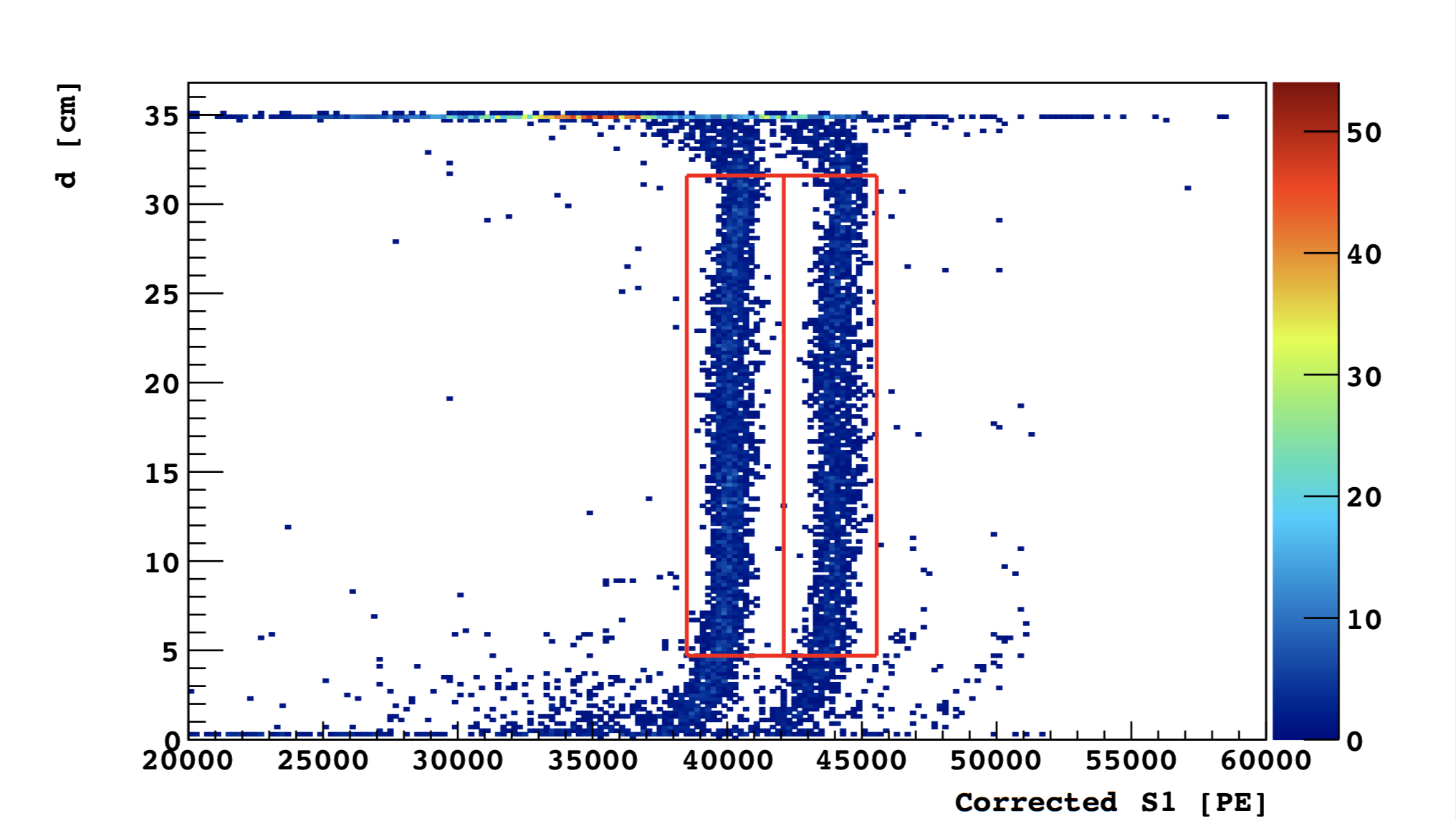}
  \caption{Plot of reconstructed depth ($d$) versus S1, corrected for saturation and depth-dependent light yield, for alpha events. The left and right bands are $^{222}$Rn and $^{218}$Po respectively.}
  \label{fig:AlphaS1tdrift}
\end{figure}

 Figure~\ref{fig:AlphaS1tdrift} shows energy versus reconstructed depth calculated from the time separation between S1 and S2 for all events with an alpha-like S1 signal \cite{ref:AlissaThesis}. The depth ($d$) is zero at the liquid-gas interface at the top of the detector, with $d=35.0$~cm at the cathode. The integral of the S1 signal, given in units of photoelectrons [PE], is approximately proportional to event energy. The main DarkSide-50 digitizers are tuned for the WIMP search range of 30 -- 200 keV$_{nr}$ (keV nuclear recoil equivalent), so alpha decays ($>$ 5 MeV) suffer from saturation effects. We correct for saturation using data from a set of low gain digitizers. S1 is also corrected for a depth-dependent light yield. This is typically done as a function of the depth reconstructed from S1 and S2. However, in order to remain consistent with other alpha studies \cite{ref:AlissaThesis, ref:ChrisThesis}, we use an alternative correction to S1 where the depth of an event is inferred from the difference in light seen between the top (S1$_\text{t}$) and bottom (S1$_\text{b}$) PMT arrays, referred to as the top-bottom asymmetry (TBA):
\begin{equation}
TBA = \frac{S1_\text{t} - S1_\text{b}}{S1_\text{t} + S1_\text{b}}.
\end{equation}
\par The two vertical bands in Fig. \ref{fig:AlphaS1tdrift} represent $^{222}$Rn on the left and $^{218}$Po on the right. More details of the identification of specific alpha emitting isotopes in the DarkSide-50 detector can be found in Ref. \cite{ref:AlissaThesis}. The bands are straight in the middle, but bend to fewer PE at the top and bottom of the detector. This is due to our TBA correction to S1. Near either array of PMTs, the variable TBA depends heavily on whether an event is located directly over a PMT face or over the Teflon reflector between PMTs. To avoid the breakdown of our correction we restrict our search to the region where the bands are straight and well separated, 
corresponding to a depth in the LAr of $d = $~[4.7, 31.6]~cm from the liquid-gas interface. 
A full description of the selection criteria for the $^{222}$Rn and $^{218}$Po decays identified for this work can be found in Chapter 5 of Ref. \cite{ref:AlissaThesis}.

\begin{figure}
\centering
  \includegraphics[width=\linewidth]{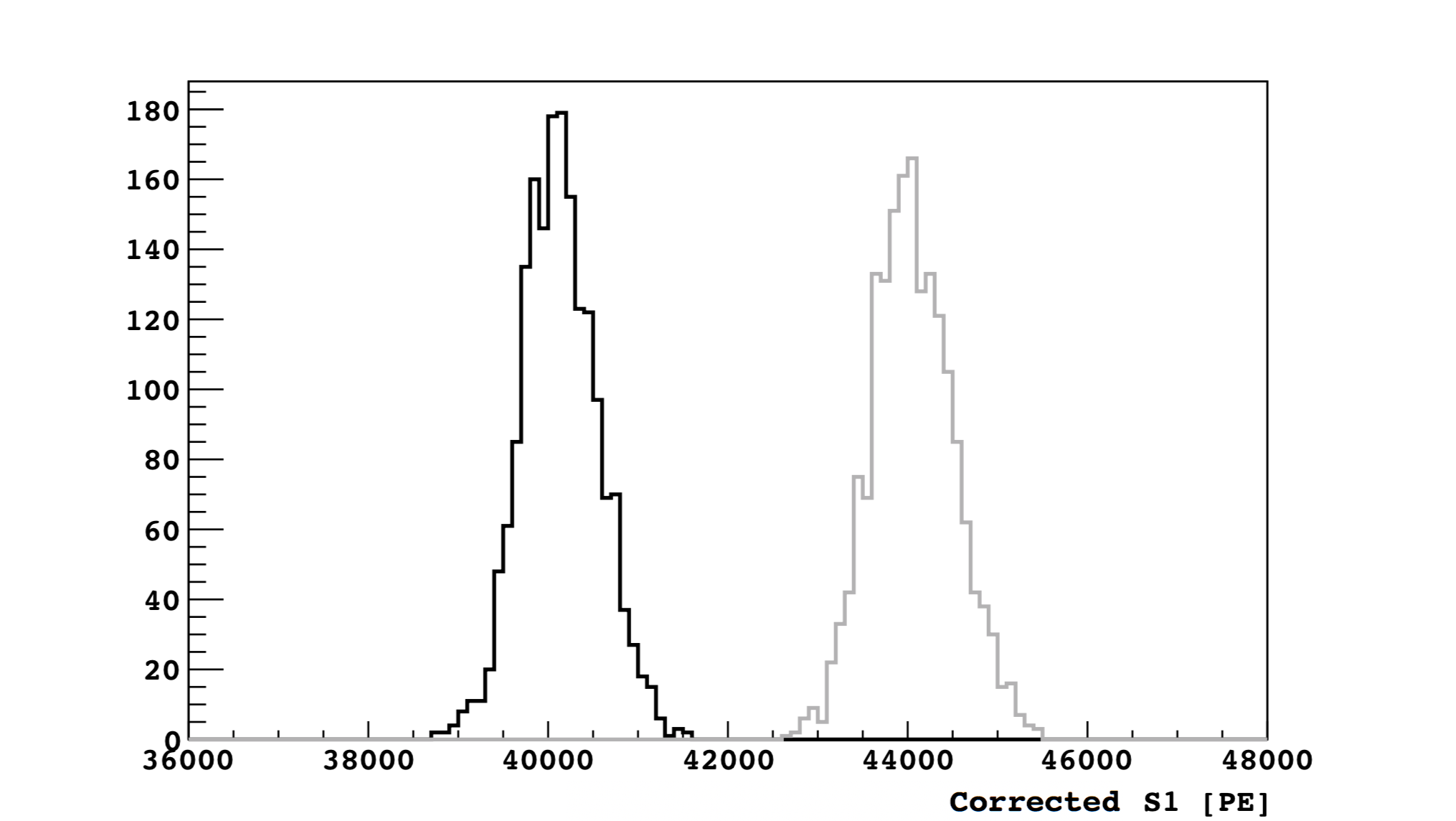}
  \caption{S1 spectra, corrected for saturation and depth-dependent light yield, for the $^{222}$Rn (black) and $^{218}$Po (gray) events in identified radon-polonium coincidences within the restricted analysis volume.}
  \label{fig:RnPoDataS1Spectra}
\end{figure}

Our radon-polonium coincidence search algorithm associates identified $^{222}$Rn events with the next identified $^{218}$Po event. If a second $^{222}$Rn event occurs, the first $^{222}$Rn event is ignored and the pairing continues on the second event. If two $^{218}$Po events are found in a row, only the first is associated with the preceding $^{222}$Rn event.We would expect to see two $^{222}$Rn events in a row when a $^{218}$Po ion drifts out of the restricted analysis volume. Likewise, we can get two $^{218}$Po events in a row when $^{222}$Rn decays above the analysis volume and the corresponding $^{218}$Po drifts into the volume. Note that our algorithm assumes that radon-polonium coincidences can be reliably tagged by pairing the $^{222}$Rn and $^{218}$Po events that are closest together in time. Given the observed rates of (7.62 $\pm$ 0.12) $^{222}$Rn decays per day and (6.61 $\pm$ 0.11) $^{218}$Po decays per day \cite{ref:AlissaThesis}, this is reasonable. We are using a reduced volume, so we expect roughly 77\% of the quoted rates. A search for radon-polonium coincidences in the 532 live-day dataset results in 1795 identified radon-polonium pairs.
The energy separation is very clear, as evidenced by Fig.~\ref{fig:RnPoDataS1Spectra}, which shows the S1 spectra for the constituent events of each identified radon-polonium coincidence, with the $^{222}$Rn decay appearing in black and the $^{218}$Po decay in gray. 

\section{Measuring the Ion Fraction and Mobility from a Drift Velocity Spectrum}
\label{sec:RnPoDataVel}

\par
\begin{figure}
\centering
  \includegraphics[width=\linewidth]{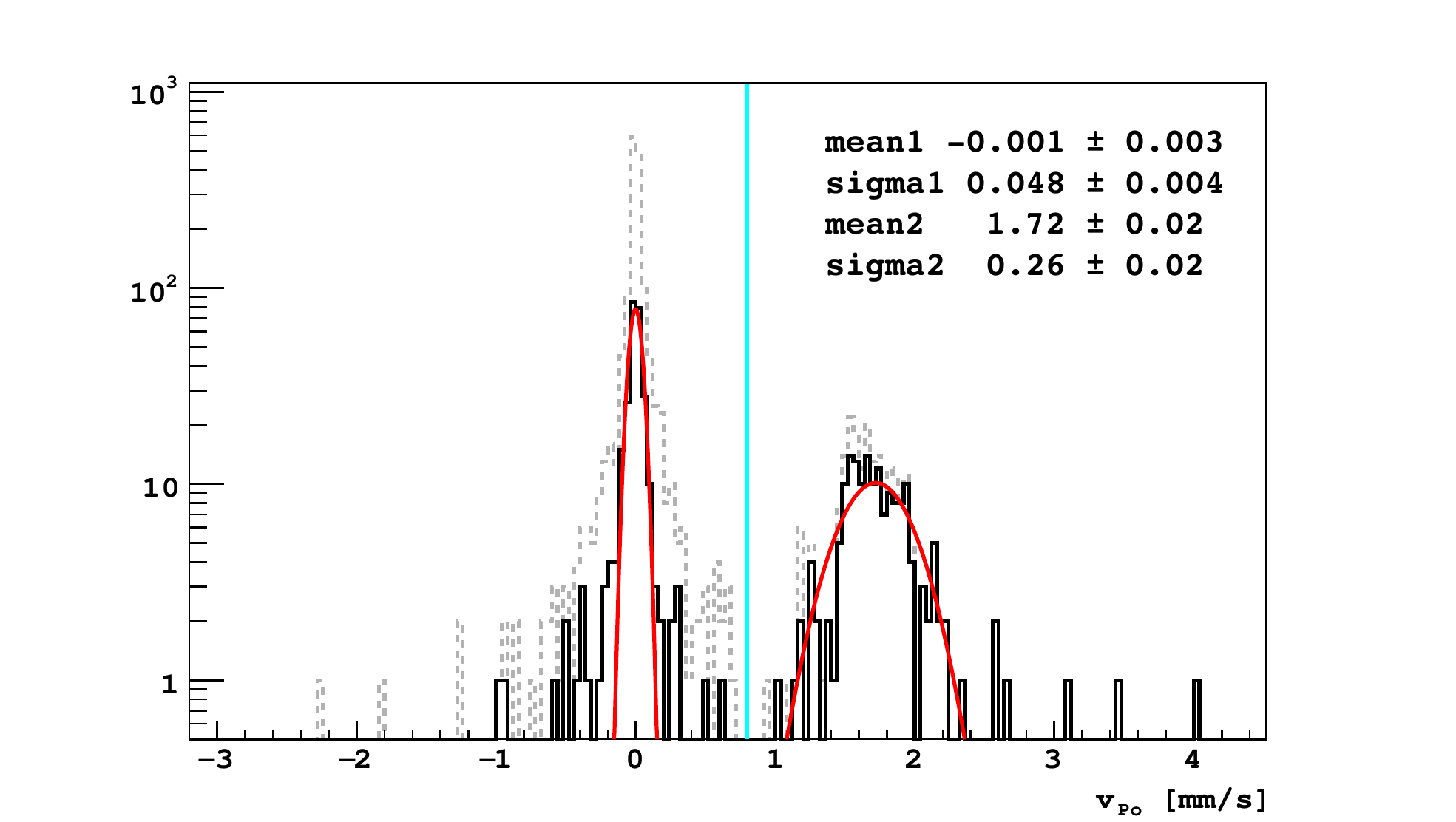}
  \caption{Plot of the average velocity of $^{218}$Po for all radon-polonium pairs(gray, dashed line) and those pairs passing Eq. \ref{eq:AddReq} (black, solid line).}
  \label{fig:RnPoDataVel}
\end{figure}

 We can measure the fraction of ionized $^{218}$Po and the mobility of $^{218}$Po ions by examining the observed spatial separation of the constituent decays in the identified radon-polonium pairs. Those pairs containing charged $^{218}$Po are more likely to be excluded from the data sample because the $^{218}$Po ion can drift out of the analysis volume. For these measurements, specifically the measurement of the charged daughter fraction ($f$), we need to apply an additional requirement to the identified radon-polonium pairs to ensure a balanced selection. To this end we require:
\begin{equation}
\label{eq:AddReq}
d_{Rn} < d_{max} - v_{Po}^{hi}\cdot t_{decay}
\end{equation}
where $d_{Rn}$ is the observed depth of the $^{222}$Rn decay, $d_{max} = 31.6$~cm is the depth of the bottom of the analysis volume (Fig. \ref{fig:AlphaS1tdrift}), $v_{Po}^{hi}$ is a chosen upper bound on the velocity of $^{218}$Po ions, and $t_{decay}$ is the observed $^{218}$Po decay time measured from the time of the observed $^{222}$Rn decay. The value of $v_{Po}^{hi}$ is defined as $v_{Po}^{avg}$ + 2$\sigma$, where $v_{Po}^{avg}$ and $\sigma$ are extracted from a Gaussian fit to the higher velocity population in an initial velocity plot filled with all radon-polonium pairs, shown as a gray dashed line in Fig. \ref{fig:RnPoDataVel}. We measure $v_{Po}^{hi}$ = 2.24~mm/s.
\par Eq.~\ref{eq:AddReq} requires that the starting position of the $^{222}$Rn decay be high enough in the detector that, given the observed $^{218}$Po decay time, the corresponding $^{218}$Po decay would still be contained within the analysis volume if it was charged and moving at $v_{Po}^{hi}$. This additional requirement effectively removes radon-polonium pairs with neutral $^{218}$Po in the same proportion as $^{218}$Po$^+$ exits the analysis volume.

\par Fig.~\ref{fig:RnPoDataVel} shows the velocity spectrum for all radon-polonium pairs (gray, dashed line), and those pairs passing the additional depth requirement (solid black line). The additional requirement reduces the sample from 1795 to 461 identified pairs. The ion velocity is calculated from the data by $v_{Po} = (d_{Po} - d_{Rn})/t_{decay}$. There are two populations in this plot, one at zero velocity and another at $\sim$1.7~mm/s. This indicates that there is only one charged species of $^{218}$Po ion observed in these $^{222}$Rn decays. This is consistent with expectations given that the first and second ionization potentials of polonium are 8.42~eV and 19.28~eV, respectively, and the ionization potential of LAr is 15.76~eV \cite{ref:ArIE}. 
\par A Gaussian fit to the higher velocity population in the spectrum of radon-polonium pairs passing the additional requirement (black) yields an ion velocity of $v_{ion}$ = (1.72 $\pm$ 0.02)~mm/s. 
The measured ion velocity is related to the ion mobility by:
\begin{equation}
\label{eq:MobVel}
v_{ion} = \mu_{ion}E
\end{equation}
where $\mu_{ion}$ is the ion mobility [10$^{-4}$cm$^2$/(V s)] and E is the drift electric field strength [V/cm] (200~V/cm in DarkSide-50).
Using Eq.~\ref{eq:MobVel}, our measured ion velocity corresponds to an ion mobility of (8.6 $\pm$ 0.1) $\times 10^{-4}$~cm$^2$/(V~s). The spread of the population centered around zero allows us to constrain the motion of $^{218}$Po due to LAr motion or diffusion along the axis of the TPC. The data is well-fit by a Gaussian with 2$\sigma <$ 0.1~mm/s. The error on the measured ion velocity in this analysis does not include the effects of LAr motion or diffusion, which are thought to influence the spread of the distribution, not the mean, as evidenced by the symmetry of the peak in the population centered on zero.
The two velocity populations are well-separated. The fraction of $^{218}$Po that is charged as a result of the $^{222}$Rn decay is calculated by counting the number of identified coincidences falling above and below 0.8~mm/s (indicated by the line at 0.8~mm/s indicated in \ref{fig:RnPoDataVel}), yielding 0.37 $\pm$ 0.03.

\par
\begin{figure}
\centering
  \includegraphics[width=\linewidth]{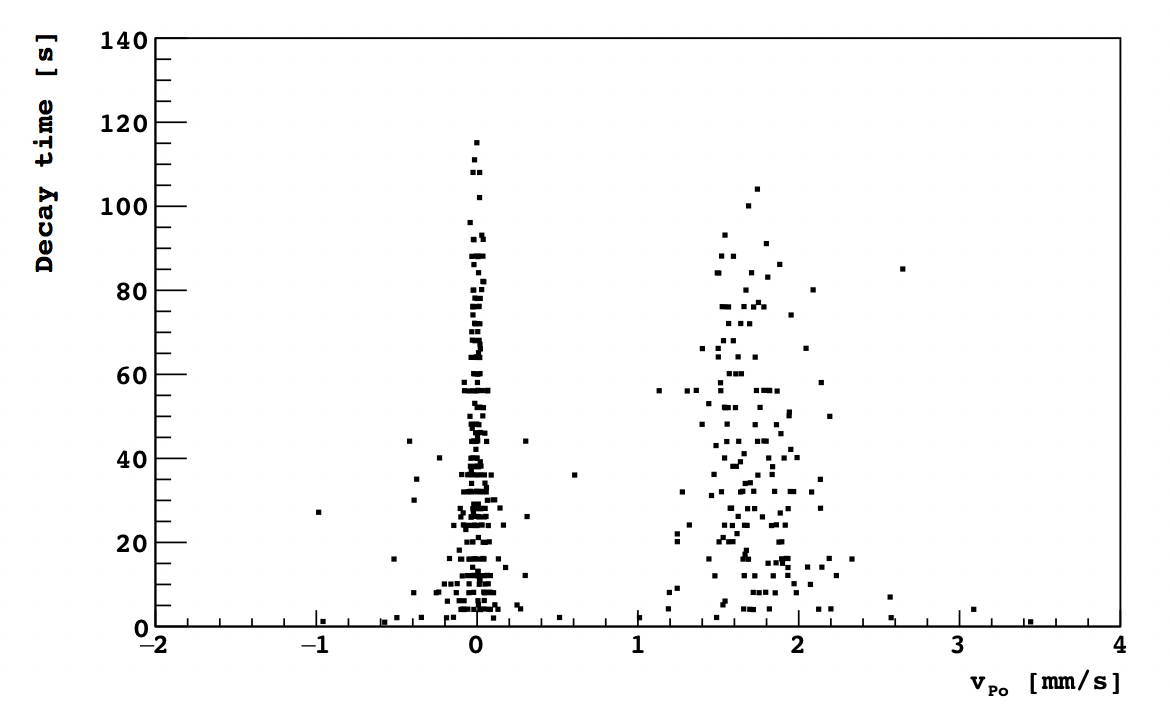}
  \caption{Plot of the average velocity of $^{218}$Po versus decay time for radon-polonium pairs selected as described in the text.}
  \label{fig:RnPoDataVeltdrift}
\end{figure}

 Note that Fig.~\ref{fig:RnPoDataVel} represents the average velocity of $^{218}$Po over the full drift time since we simply divide the total position difference by the total time elapsed between decays. If the $^{218}$Po ion drift velocity changes as a function of time, due to impurities for example, that would not be captured in this plot. In EXO-200, a drift time-dependent ion velocity was observed, and was ascribed to the presence of impurities in the LXe \cite{ref:EXOMobility}. To check the stability of the ion velocity in DarkSide-50 as a function of the time the ion spends drifting, $v_{Po}$ versus decay time is plotted in Fig.~\ref{fig:RnPoDataVeltdrift}. A Pearson correlation coefficient of -0.15 is calculated for the population centered around 1.72~mm/s, indicating a very weak correlation. We conclude that the velocity of $^{218}$Po ions does not have an obvious dependence on the ion drift time. 

\section{Modeling the Decay Time Spectrum}
\label{sec:RnPoSim}

\par When charged $^{218}$Po daughters drift out of our analysis volume, the corresponding radon-polonium pair does not pass our selection criteria. The probability for a $^{218}$Po$^+$ ion to escape our analysis volume changes as a function of the ion drift time (its decay time), manifesting as a deviation of the observed $^{218}$Po decay time spectrum away from a pure exponential. The shape of this spectrum can be described analytically and fit to the decay time spectrum in data to extract parameters of interest. This is done in Sec. \ref{sec:RnPoTDecayFits}. To keep our model general we will initially consider an analysis volume that extends all the way to the cathode. For the purposes of the derivation we will define several useful parameters:
\begin{itemize}
\item $\tau$; the lifetime of $^{218}$Po.
\item $t_{max}$; the maximum drift time possible for $^{218}$Po ions, a function of ion mobility and the detector geometry. 
\item $f$; the fraction of charged $^{218}$Po.
\item $P_c$; the probability to observe a decay on the cathode.
\end{itemize}
\par We assume that the $^{218}$Po daughters come from a spatially uniform distribution of $^{222}$Rn. Then, for a $^{218}$Po decay time $t$, the fraction of $^{218}$Po ions that reach the cathode is $\frac{t}{t_{max}}$ for  $t < t_{max}$, while for $t \geq t_{max}$ all $^{218}$Po ions have already reached the cathode. 
We split the radon-polonium pairs into those with neutral and charged $^{218}$Po. For the neutral case, the resulting decay time spectrum is simply
\begin{equation}
\label{eq:NeutralDaughters}
F_{neutral}(t) = A (1 - f) e^\frac{-t}{\tau}
\end{equation}
where $A$ is a normalization, and $t$ is the $^{218}$Po decay time. For the charged case, the resulting decay time spectrum is
\begin{equation}
\label{eq:ChargedDaughters}
F_{charged}(t) = \begin{cases}
  A f \left[ \left( 1 - \frac{t}{t_{max}} \right) + P_c \frac{t}{t_{max}} \right] e^\frac{-t}{\tau} & t < t_{max}\\
  A f P_c e^\frac{-t}{\tau} & t \geq t_{max}
  \end{cases}
\end{equation}
The full expression for the analytical description of observed radon-polonium decays in a TPC is the sum of Eqns.~\ref{eq:NeutralDaughters} and~\ref{eq:ChargedDaughters}:
 \begin{equation}
\label{eq:CathodeLossFunc}
F(t) = F_{neutral}(t) + F_{charged}(t) = \begin{cases}
  A e^\frac{-t}{\tau} \left[1 - f (1 - P_c) \frac{t}{t_{max}} \right] & t < t_{max}\\
  A e^\frac{-t}{\tau} \left[ 1 - f \left( 1 - P_c \right) \right] & t \geq t_{max}
  \end{cases}
\end{equation}
\par Our restricted analysis volume requires two modifications to Eq. \ref{eq:CathodeLossFunc}. The restricted volume decreases the maximum drift time for ions across the volume, shortening $t_{max}$. More importantly, once a $^{218}$Po ion leaves the volume it has zero probability to be observed, so $P_c = 0$. The function then becomes
 \begin{equation}
\label{eq:ModCathodeLossFunc}
F_{Mod}(t) = const + \begin{cases}
  A e^\frac{-t}{\tau} \left(1 - f \frac{t}{t_{max}} \right) & t < t_{max}\\
  A e^\frac{-t}{\tau} \left( 1 - f \right) & t \geq t_{max}
  \end{cases},
\end{equation}
where we have added a constant to handle possible background in a fit to our data.

\par In order to illustrate the modeled effect of detection efficiency and restricted analysis volumes on the observed decay time spectrum, a toy Monte Carlo is performed. We simulate radon-polonium pairs with $\tau$ = 268.2~s \cite{ref:Tau}, a charged daughter fraction of $f$ = 0.37, and ion mobility of $\mu_{ion}$ = 8.6 $\times 10^{-4}$~cm$^2$/(V s), where we have used the values measured from DarkSide-50 data in Sec. \ref{sec:RnPoDataVel}. The simulation mimics the DarkSide-50 geometry, with a full drift length of 35.0~cm. Decays that occur on the cathode are detected with a 50\% efficiency, as an alpha particle emitted into the cathode will deposit no energy in the LAr. We generate 10$^8$ radon-polonium coincidences, with the position of the $^{222}$Rn decay randomly distributed throughout the simulated volume. Uncharged $^{218}$Po remain and decay at their initial position, but $^{218}$Po$^+$ ions drift towards the cathode at the speed defined by $\mu_{ion}$.

\par
\begin{figure}
\centering
  \includegraphics[width=\linewidth]{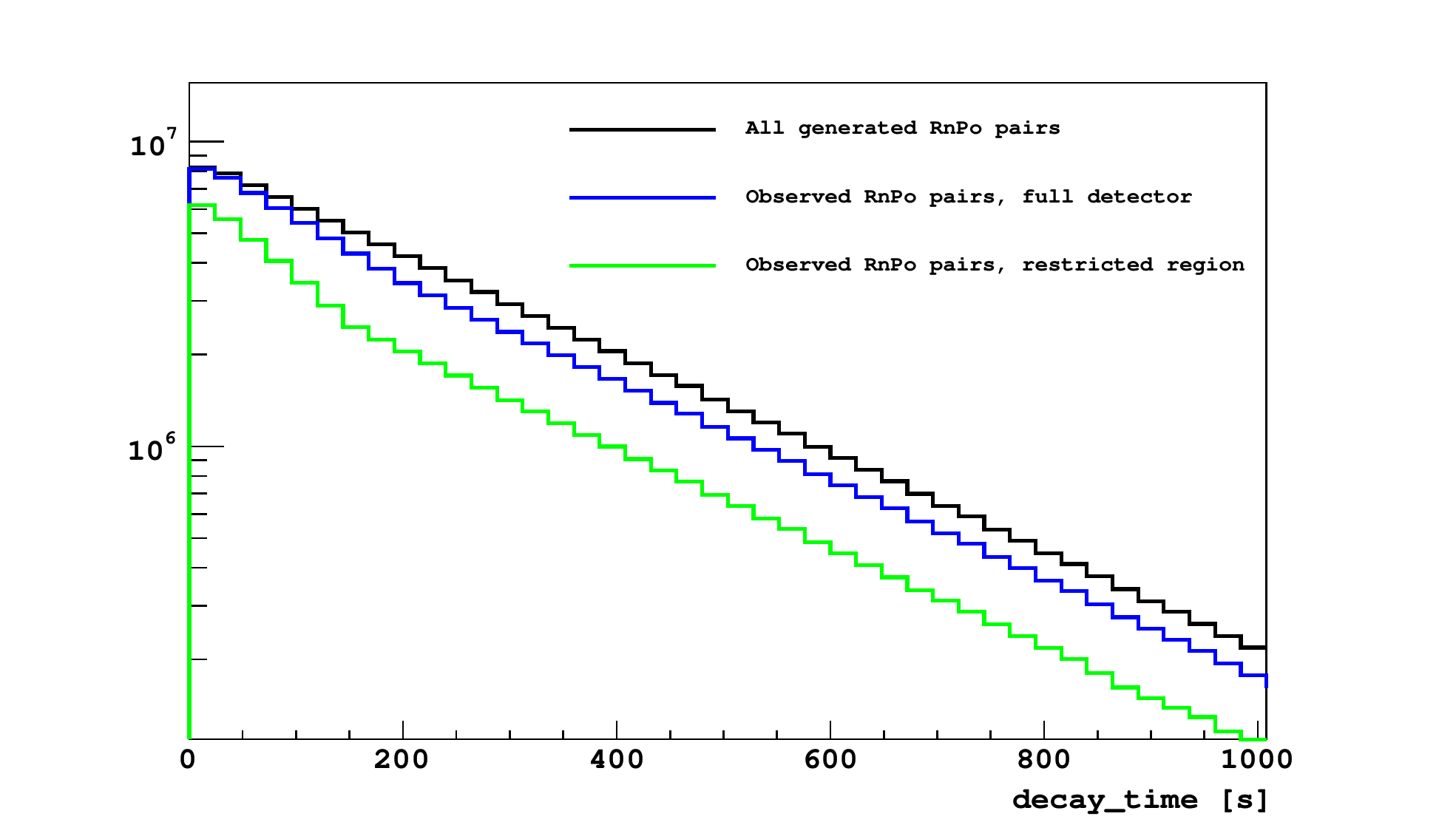}
  \caption{Toy monte carlo results for the decay time of simulated radon-polonium coincidences.}
  \label{fig:RnPoSimTDecay}
\end{figure}

 Figure~\ref{fig:RnPoSimTDecay} shows the simulated decay time spectrum, color coded for various event selections. 
The spectrum of all generated radon-polonium pairs (black) exhibits pure exponential decay. The spectrum of observed radon-polonium pairs in the full volume (blue) exhibits a suppression at longer decay times and a kink around 200~s. The position of this feature is set by the full drift length for ions.\footnote{Given the input values of the simulation: the nominal 200~V/cm drift field in DarkSide-50, a full drift length of 35.0~cm, and an ion mobility of 8.6 $\times 10^{-4}$~cm$^2$/(V s), the corresponding maximum ion drift time is 203~s.} The spectrum of observed decays for a restricted analysis region equivalent to the one used in this work (green) exhibits a stronger suppression and kink around 160~s. The position of the kink changes due to the shorter drift length across the reduced analysis volume, 27.0~cm versus 35.0~cm.

\section{Fit to the Decay Time Spectrum}
\label{sec:RnPoTDecayFits}

\par
\begin{figure}
\centering
  \includegraphics[width=\linewidth]{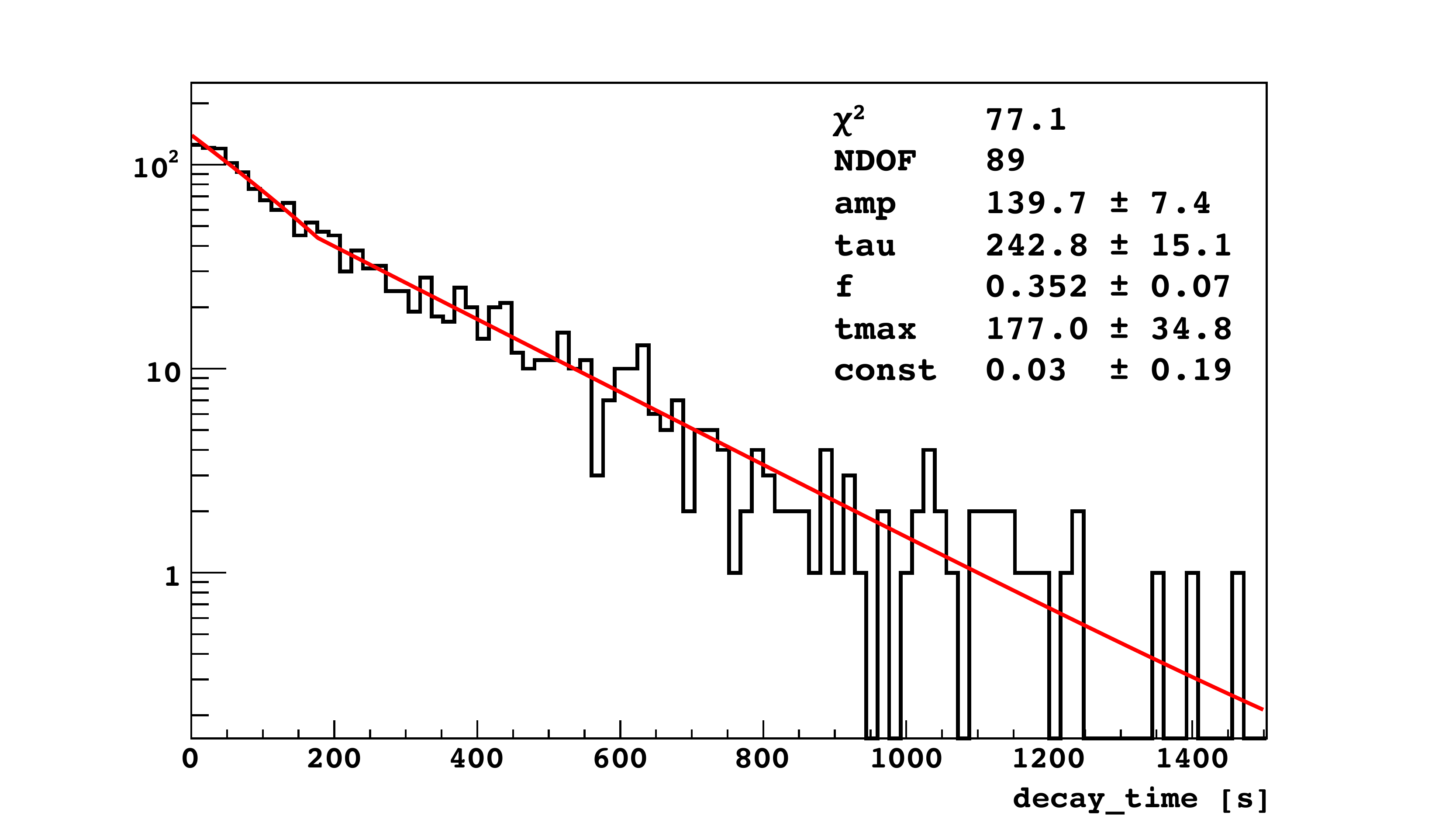}
  \caption{Decay time spectrum for radon-polonium pairs identified in 532 live-days of DarkSide-50 data.}
  \label{fig:RnPoDataTDecay}
\end{figure}

 The decay time spectrum for identified radon-polonium pairs in DarkSide-50 data is shown in Fig.~\ref{fig:RnPoDataTDecay}. A binned likelihood fit of Eq.~\ref{eq:ModCathodeLossFunc} to the data yields $\tau$ = (242.8 $\pm$ 15.1)~s, $f$ = (0.35 $\pm$ 0.07), and $t_{max}$ = (177.0 $\pm$ 34.8)~s. We can relate the maximum drift time for ions to the ion velocity by:
\begin{equation}
    v_{ion} = \frac{D}{t_{max}}
    \label{eq:velTMax}
\end{equation}
where $D=27.0 \pm 0.3$~cm is the maximum drift length across the reduced analysis volume at the standard field configuration (E = 200 V/cm). This definition of $t_{max}$ is precise for an ion drifting at exactly the mean value of $v_{ion}$. However, as observed in Fig. \ref{fig:RnPoDataVel}, there is a spread in ion velocity. When relating $v_{ion}$ and $t_{max}$, we consider $t_{max}$ to be the mean value of a distribution.

\par The value of $\tau$ extracted from the fit is within 2$\sigma$ of the value in Ref. \cite{ref:Tau}. 
Using Equation \ref{eq:velTMax}, a fitted $t_{max}$ value of (177.0 $\pm$ 34.8)~s corresponds to an ion drift velocity of (1.52 $\pm$ 0.30)~mm/s and an ion mobility of (7.6 $\pm$ 1.5)$\times10^{-4}$~cm$^2$/(V s). Likewise, the measured $v_{ion} = 1.72 \pm 0.02$~mm/s from Sec. \ref{sec:RnPoDataVel} corresponds to $t_{max}$ = 157.0 $\pm$ 2.5 s. The small value of the constant returned by the fit, compatible with zero, confirms the cleanliness of the radon-polonium coincidence sample. The values of $f$ and $t_{max}$ from the decay time spectrum fit are within one standard deviation of the values extracted from Fig. \ref{fig:RnPoDataVel}. 
\par The fact that the reduced $\chi^2$ is less than one is expected, given that the model is an over-fit to the data and contains multiple correlated parameters. The Pearson correlation coefficients for the parameters in the fit are summarized in Table~\ref{tab:correlations}.
A number of the parameters are moderately to highly correlated. From this we conclude that the decay time spectrum fit is not sensitive enough to make an independent statement on the values of those parameters. However, it provides a consistency check on the values extracted by more sensitive means in Sec.~\ref{sec:RnPoDataVel}.

\begin{table}
\centering
\begin{tabular}{c|ccccc}
\hline
   & amp & tau & f & $t_{max}$ & const\\
\hline
  amp & 1 & -- & -- & -- & --\\
  tau & -0.03 & 1 & -- & -- & --\\
  f & 0.46 & 0.79 & 1 & -- & --\\
  $t_{max}$ & -0.43 & 0.39 & 0.25 & 1 & --\\
  const & 0.01 & -0.74 & -0.49 & -0.21 & 1\\
  \end{tabular}
 \caption{Table of Pearson correlation coefficients for the fit parameters from Fig.~\ref{fig:RnPoDataTDecay}. Redundant coefficients are omitted.}
\label{tab:correlations}
\end{table}

\section{Results and Discussion}
\label{sec:RnPoResults}
A summary of the measurements from Sections~\ref{sec:RnPoTDecayFits} and~\ref{sec:RnPoDataVel} are shown in Tab.~\ref{tab:RnPoSummary}, along with other measurements from the literature. The values from the two measurements conducted in this study (the analysis of $^{218}$Po velocity and fits to its decay time spectrum) are consistent. The measurements from this study are also within errors of the analysis performed in Ref. \cite{ref:ChrisThesis}. 
\par The ion mobility in liquid argon is higher than in liquid xenon \cite{ref:EXOMobility}, with the mobility differing by a factor of 2.2 $\pm$ 0.04. It is not unexpected that the measured mobilities in Ar and Xe would differ, as the two liquids have different densities. In fact, the relative density of xenon to argon is 2.942 [$\frac{\text{g}}{\text{cm}^3}$]/1.395 [$\frac{\text{g}}{\text{cm}^3}$] = 2.1 \cite{ref:EncAirLiquide}. The difference in the measured mobilities is consistent with an inversely proportional relationship between density and the mobility of ions in the noble liquids.
\par The fraction of charged $^{218}$Po produced in the $^{222}$Rn decay in LXe (0.503 $\pm$ 0.03) is higher than in LAr (0.37 $\pm$ 0.03) by a factor of 1.35 $\pm$ 0.17 (recall the relative density of xenon to argon is 2.1). 
The relationship between the ion fraction and liquid density does not appear to be linear.

\section{Conclusions}
The charged daughter fraction of $^{218}$Po and the mobility of $^{218}$Po$^{+}$ ions was measured by two independent methods using identified radon-polonium coincidences in DarkSide-50 data. The two measurements presented in this study, a fit to the observed decay time spectrum and examination of the observed velocity spectrum, are consistent. The fraction of charged $^{218}$Po produced in the $^{222}$Rn decay in LAr is measured to be (0.37 $\pm$ 0.03) and the observed mobility of $^{218}$Po$^+$ ions is (8.6~$\pm$~0.1)~$\times$~10$^{-4}$~$\frac{\text{cm}^2}{\text{Vs}}$.

\begin{table}
\centering
\begin{tabular}{cccccc}
\hline
  \thead{\textbf{Experiment} \\\textbf{Medium}} & Ref. & \thead{$t_{max}$ \\ $[\text{s}]$} & \thead{$v_{ion}$ \\ $[\frac{\text{mm}}{\text{s}}]$} & \thead{$\mu$ \\ $[10^{-4}\frac{\text{cm}^2}{\text{Vs}}]$} & $f$\\
\hline
  \thead{\textbf{EXO-200} \\\textbf{LXe}} & \cite{ref:EXOMobility} & -- & 1.48 $\pm$ 0.01 & 3.90 $\pm$ 0.06 & 0.503 $\pm$ 0.03\\
  \thead{\textbf{DS-50} \\\textbf{LAr}} & \cite{ref:ChrisThesis} & -- & 1.53 $\pm$0.05 & 7.64 & 0.36 $\pm$ 0.05\\
  \thead{\textbf{DS-50} \\\textbf{LAr}} & \thead{This work \\($t_{decay}$ fit)} & 177.0 $\pm$ 34.8 & 1.52 $\pm$ 0.30 & 7.6 $\pm$ 1.5 & 0.35 $\pm$ 0.07\\
  \thead{\textbf{DS-50} \\\textbf{LAr}} & \thead{This work \\($v_{ion}$ plot)} &157.0 $\pm$ 2.5 & 1.72 $\pm$ 0.02 & 8.6 $\pm$ 0.1 & 0.37 $\pm$ 0.03\\
  \end{tabular}
 \caption{Table summarizing the results of radon-polonium analysis with comparisons to literature. The uncertainties represent statistical uncertainties only.}
\label{tab:RnPoSummary}
\end{table}

\acknowledgments

This work was supported by the US NSF (Grants PHY-0919363, PHY-1004072, PHY-1004054, PHY-1242585, PHY-1314483, PHY-1606912, PHY-1314507 and associated collaborative grants; grants PHY-1211308 and PHY-1455351), the Italian Istituto Nazionale di Fisica Nucleare (INFN), the US DOE (Contract Nos. DE-FG02- 91ER40671 and DE-AC02-07CH11359), the Russian RSF (Grant No 16-12-10369), and the Polish NCN (Grant UMO-2014/15/B/ST2/02561). We thank the staff of the Fermilab Particle Physics, Scientific and Core Computing Divisions for their support. We acknowledge the financial support from the UnivEarthS Labex program of Sorbonne Paris Cit\'e (ANR-10-LABX-0023 and ANR-11-IDEX-0005-02), from \~Sao Paulo Research Foundation (FAPESP) grant (2016/09084-0), and from Foundation for Polish Science (grant No. TEAM/2016-2/17). We would also like to thank Brian Mong for insightful discussions about ion mobility analyses.


\end{document}

%% file: ionMobility-frontmatter.tex


  \title{Measurement of the ion fraction and mobility of $^{218}$Po produced in $^{222}$Rn decays in liquid argon}

  
  \input{AU-50-2019.tex}
  \input{inst}


\abstract{We report measurements of the charged daughter fraction of $^{218}$Po as a result of the $^{222}$Rn alpha decay, and the mobility of $^{218}$Po$^+$ ions, using radon-polonium coincidences from the $^{238}$U chain identified in 532 live-days of DarkSide-50 WIMP-search data.  The fraction of $^{218}$Po that is charged is found to be 0.37~$\pm$~0.03 and the mobility of $^{218}$Po$^+$ is (8.6~$\pm$~0.1)~$\times$~10$^{-4}$~$\frac{\text{cm}^2}{\text{Vs}}$.}



%% file: AU-50-2019.tex
\author[a]{P.~Agnes}
\author[b]{I.F.M.~Albuquerque}
\author[c]{T.~Alexander}
\author[d]{A.K.~Alton}
\author[b]{M.~Ave}
\author[c]{H.O.~Back}
\author[e,f]{G.~Batignani}
\author[g]{K.~Biery}
\author[h]{V.~Bocci}
\author[i]{G.~Bonfini}
% ...

%% file: inst.tex
\affiliation[a]{Department of Physics, University of Houston, Houston, TX 77204, USA}
\affiliation[b]{Instituto de F\'isica, Universidade de S\~ao Paulo, S\~ao Paulo 05508-090, Brazil}
\affiliation[c]{Pacific Northwest National Laboratory, Richland, WA 99352, USA}
\affiliation[d]{Physics Department, Augustana University, Sioux Falls, SD 57197, USA}
\affiliation[e]{INFN Pisa, Pisa 56127, Italy}
\affiliation[f]{Physics Department, Universit\`a degli Studi di Pisa, Pisa 56127, Italy}
\affiliation[g]{Fermi National Accelerator Laboratory, Batavia, IL 60510, USA}
\affiliation[h]{INFN Sezione di Roma, Roma 00185, Italy}
\affiliation[i]{INFN Laboratori Nazionali del Gran Sasso, Assergi (AQ) 67100, Italy}
\affiliation[j]{INFN Cagliari, Cagliari 09042, Italy}
\affiliation[k]{Physics Department, Universit\`a degli Studi di Genova, Genova 16146, Italy}
\affiliation[l]{INFN Genova, Genova 16146, Italy}
\affiliation[m]{INFN Roma Tre, Roma 00146, Italy}
\affiliation[n]{Mathematics and Physics Department, Universit\`a degli Studi Roma Tre, Roma 00146, Italy}
\affiliation[o]{Physics Department, Universit\`a degli Studi di Cagliari, Cagliari 09042, Italy}
\affiliation[p]{Physics Department, Princeton University, Princeton, NJ 08544, USA}
\affiliation[q]{Gran Sasso Science Institute, L'Aquila 67100, Italy}
\affiliation[r]{Chemistry and Pharmacy Department, Universit\`a degli Studi di Sassari, Sassari 07100, Italy}
\affiliation[s]{INFN Laboratori Nazionali del Sud, Catania 95123, Italy}
\affiliation[t]{Physics Department, Universit\`a degli Studi ``Federico II'' di Napoli, Napoli 80126, Italy}
\affiliation[u]{INFN Napoli, Napoli 80126, Italy}
\affiliation[v]{Virginia Tech, Blacksburg, VA 24061, USA}
\affiliation[w]{Skobeltsyn Institute of Nuclear Physics, Lomonosov Moscow State University, Moscow 119234, Russia}
\affiliation[x]{Physics Department, Universit\`a degli Studi di Milano, Milano 20133, Italy}
\affiliation[y]{INFN Milano, Milano 20133, Italy}
\affiliation[z]{Physics Department, Sapienza Universit\`a di Roma, Roma 00185, Italy}
\affiliation[aa]{Saint Petersburg Nuclear Physics Institute, Gatchina 188350, Russia}
\affiliation[ab]{Amherst Center for Fundamental Interactions and Physics Department, University of Massachusetts, Amherst, MA 01003, USA}
\affiliation[ac]{Department of Physics and Astronomy, University of Hawai'i, Honolulu, HI 96822, USA}
\affiliation[ad]{Joint Institute for Nuclear Research, Dubna 141980, Russia}
\affiliation[ae]{APC, Universit\'e Paris Diderot, CNRS/IN2P3, CEA/Irfu, Obs de Paris, USPC, Paris 75205, France}
\affiliation[af]{Museo della fisica e Centro studi e Ricerche Enrico Fermi, Roma 00184, Italy}
\affiliation[ag]{LPNHE, CNRS/IN2P3, Sorbonne Universit\'e, Universit\'e Paris Diderot, Paris 75252, France}
\affiliation[ah]{Physics Department, Temple University, Philadelphia, PA 19122, USA}
\affiliation[ai]{National Research Centre Kurchatov Institute, Moscow 123182, Russia}
\affiliation[aj]{National Research Nuclear University MEPhI, Moscow 115409, Russia}
\affiliation[ak]{Institute of High Energy Physics, Beijing 100049, China}
\affiliation[al]{Engineering and Architecture Faculty, Universit\`a di Enna Kore, Enna 94100, Italy}
\affiliation[am]{Department of Physics, University of California, Davis, CA 95616, USA}
\affiliation[an]{School of Natural Sciences, Black Hills State University, Spearfish, SD 57799, USA}
\affiliation[ao]{Radiation Physics Laboratory, Belgorod National Research University, Belgorod 308007, Russia}
\affiliation[ap]{Physics Institute, Universidade Estadual de Campinas, Campinas 13083, Brazil}
\affiliation[aq]{Chemistry, Biology and Biotechnology Department, Universit\`a degli Studi di Perugia, Perugia 06123, Italy}
\affiliation[ar]{INFN Perugia, Perugia 06123, Italy}
\affiliation[as]{M. Smoluchowski Institute of Physics, Jagiellonian University, 30-348 Krakow, Poland}
\affiliation[at]{Physics and Astronomy Department, University of California, Los Angeles, CA 90095, USA}

%% file: ionMobility.bbl
\begin{thebibliography}{99}

\bibitem{ref:HeavyIons}
A. J. Howard and W. P. Strange, \emph{Heavy-ion migration through argon and helium in weak electric fields}, \href{https://doi.org/10.1063/1.348822}{\emph{J. Appl. Phys.} {\bf 69} 6248 (1991)}. 

\bibitem{ref:LUXRn}
A. Bradley, et. al. (LUX Collaboration), \emph{Radon-related Backgrounds in the LUX Dark Matter Search}, \href{https://doi.org/10.1016/j.phpro.2014.12.067}{\emph{Phys. Proc.} {\bf{61}} 658-665 (2015)}. 

\bibitem{ref:XENONRn}
E. Aprile, et. al. (XENON Collaboration), \emph{Intrinsic backgrounds from Rn and Kr in the XENON100 experiment}, \href{https://doi.org/10.1140/epjc/s10052-018-5565-y}{\emph{Eur. Phys. J. C} {\bf{78}} 132 (2018)}. 

\bibitem{ref:nEXO}
J. B. Albert, et. al. (nEXO Collaboration), \emph{Sensitivity and discovery potential of the proposed nEXO experiment to neutrinoless double-$\beta$ decay}, \href{https://doi.org/10.1103/PhysRevC.97.065503}{\emph{Phys. Rev. C} {\bf 97} 065503 (2018)}. 

\bibitem{ref:GERDA}
M. Agostini, et. al. (GERDA Collaboration), \emph{The background in the 0$\nu\beta\beta$ experiment GERDA}, \href{https://doi.org/10.1140/epjc/s10052-014-2764-z}{\emph{Eur. Phys. J. C} {\bf{74}} 2764 (2014)}. 

\bibitem{ref:RnDaughters}
K. Pelczar, N. Frodyma, and M. Wojcik, \emph{Short-lived Rn-222 daughters in cryogenic liquids}, \href{https://doi.org/10.1063/1.4818109}{\emph{AIP Conference Proceedings} {\bf 1549} 205 (2013)}. 

\bibitem{ref:BaTagging}
C. Chambers, et. al. (nEXO Collaboration), \emph{Imaging individual barium atoms in solid xenon for barium tagging in nEXO}, \href{https://www.nature.com/articles/s41586-019-1169-4}{\emph{Nature} {\bf{569}} 203-207 (2019)}. 

\bibitem{ref:EXOMobility}
J. B. Albert, et. al. (EXO-200 Collaboration), \emph{Measurements of the ion fraction and mobility of and decay products in liquid xenon using the EXO-200 detector}, \href{https://doi.org/10.1103/PhysRevC.92.045504}{\emph{Phys. Rev. C} {\bf 92} 045504 (2015)}. 

\bibitem{ref:BaClusterIons}
E. Bainglass, et. al., \emph{Mobility and clustering of barium ions and dications in high-pressure xenon gas}, \href{https://journals.aps.org/pra/pdf/10.1103/PhysRevA.97.062509}{\emph{Phys. Rev. A }{\bf 97} 062509 (2018)}.

\bibitem{ref:ArClusterIons}
Y. Kalkan, et. al., \emph{Cluster ions in gas-based detectors}, \href{https://iopscience.iop.org/article/10.1088/1748-0221/10/07/P07004/pdf}{\emph{JINST} {\bf 10} P07004 (2015)}

\bibitem{ref:SBN}
M. Antonello, et. al. (MicroBooNE and LAr1-ND and ICARUS-WA104 Collaborations), \emph{A Proposal for a Three Detector Short-Baseline Neutrino Oscillation Program in the Fermilab Booster Neutrino Beam}, \href{http://arxiv.org/abs/arXiv:1503.01520}{{\bf arXiv:1503.01520} (2015)}. 

\bibitem{ref:DUNE}
R. Acciarri, et. al. (DUNE Collaboration), \emph{Long-Baseline Neutrino Facility (LBNF) and Deep Underground Neutrino Experiment (DUNE) : Conceptual Design Report, Volume 1: The LBNF and DUNE Projects}, \href{http://arxiv.org/abs/arXiv:1601.05823}{{\bf arXiv:1601.05471} (2016)}. 

\bibitem{ref:50d}
P. Agnes, et. al. (DarkSide Collaboration), \emph{First results from the DarkSide-50 dark matter experiment at Laboratori Nazionali del Gran Sasso}, \href{https://doi.org/10.1016/j.physletb.2015.03.012}{\emph{Phys. Lett. B} {\bf 743} 456-466 (2015)}. 

\bibitem{ref:70d}
P. Agnes, et. al. (DarkSide Collaboration), \emph{Results from the first use of low radioactivity argon in a dark matter search}, \href{https://doi.org/10.1103/PhysRevD.93.081101}{\emph{Phys. Rev. D} {\bf 93} 081101(R) (2017)}. 

\bibitem{ref:532d}
P. Agnes, et. al. (DarkSide Collaboration), \emph{DarkSide-50 532-day Dark Matter Search with Low-Radioactivity Argon}, \href{https://doi.org/10.1103/PhysRevD.98.102006}{\emph{Phys. Rev. D} {\bf{98}} 102006 (2018)}. 

\bibitem{ref:S2Only}
P. Agnes, et. al. (DarkSide Collaboration), \emph{Low-Mass Dark Matter Search with the DarkSide-50 Experiment}, \href{https://doi.org/10.1103/PhysRevLett.121.081307}{\emph{Phys. Rev. Lett.} {\bf{121}} 081307 (2018)}. 

\bibitem{ref:Tau}
G. Audi, F. G. Kondev, M. Wang, W. J. Huang, and S. Naimi, \emph{The NUBASE2016 evaluation of nuclear properties}, \href{https://doi.org/10.1088/1674-1137/41/3/030001}{\emph{Chin. Phys. C} {\bf{41}} 030001 (2017) }. 


\bibitem{ref:AlissaThesis}
A. Monte, \emph{Alpha Radiation Studies and Related Backgrounds in the DarkSide-50 Detector}, PhD thesis, University of Massachusetts Amherst (2018), URL \href{https://doi.org/10.2172/1497091}{https://doi.org/10.2172/1497091}. 

\bibitem{ref:ChrisThesis}
C. Stanford, \emph{Alphas and Surface Backgrounds in Liquid Argon Dark Matter Detectors}, PhD thesis, Princeton University (2017), URL \href{https://physics.princeton.edu/archives/theses/lib/upload/christopher_stanford.pdf}{https://physics.princeton.edu/archives/theses/lib/upload/christopher$\_$stanford.pdf}. 


\bibitem{ref:ArIE}
Sharon G. Lias, \emph{"Ionization Energy Evaluation" in NIST Chemistry WebBook}, \href{https://webbook.nist.gov/cgi/inchi?ID=C7440371&Mask=20}{\emph{NIST Standard Reference Database Number} {\bf{69}} (retrieved February 10, 2019)}. 







\bibitem{ref:EncAirLiquide}
Encyclopedia Air Liquide, Xe and Ar Comparison, \href{https://encyclopedia.airliquide.com/compare-tool/66/6}{\emph{https://encyclopedia.airliquide.com/compare-tool/66/6}} (2018).







\end{thebibliography}
